# ELECTROMAGNETIC SHIELDING WITH POLYPYRROLE-COATED FABRICS

J. Avloni[(1)], L. Florio[(2)], A.R. Henn[(3)], R. Lau[(1)], M. Ouyang[(1)], and A. Sparavigna[(4)]

[(1)] *Eeonyx Corporation, 750 Belmont Way, Pinole, CA 94564, USA*
[(2)] *Laboratorio Tessili Innovativi, ITIS "Q. Sella", Via Rosselli 2 , 13900 Biella, Italy*
[(3)] *Marktek Inc., 13621 Riverway Dr., Suite H, Chesterfield, MO 63017, USA*
[(4)] *Physics Department, Politecnico di Torino, C.so Duca degli Abruzzi 24, 10129 Torino, Italy*

Several shielding applications, to protect human health and electronic devices against dangerous effects of electromagnetic radiation, require solutions that fabrics can suitably fulfil. Here, we will investigate the electromagnetic interference shielding effectiveness of polypyrrole-coated polyester textiles, in the frequency range 100-1000 MHz. Insertion losses for several conductive fabrics with different surface resistivity ranging from $40\,\Omega$ till the very low value of $3\,\Omega$ were evaluated with a dual-tem cell. Correlations between the shielding effectiveness and the conductivity of composites are also discussed.

## I. INTRODUCTION

There is no doubt that the best materials for electromagnetic shields possess both high conductivity and high permeability and that shielding devices based on the use of metals are the best ones. However apart from military applications, metals are being increasingly replaced by thermoplastics for housing commercial equipments, due to flexibility, light weight and low cost. Metallized thermoplastic materials are now commonly used for shielding elements. Among these materials, several commercial metallized fabrics are also available. Textiles are also suitable to provide protective clothing for people exposed to high frequency electromagnetic fields, to fulfil safety requirements in the field of non-ionizing radiation.

To impart shielding properties to textiles, metallizing fabrics is an approach suitable for industrial scale processes, where textile screens are covered with metal, mainly by chemical methods [1-3]. Another possibility results in incorporating electrically conductive fillers, in the form of fibers injected in to synthetic resins during the moulding stage [4,5]. Innovative materials are intrinsically conducting polymers, such as polypyrrole (PPy) and polyaniline, as good materials to obtain an economical coating system for fabrics with natural or synthetic fibres [6].

In this paper we will discuss attenuation of electromagnetic waves with polypyrrole-coated polyester fabrics, prepared with a modified formulation giving superior stability and conductivity. Shielding effectiveness evaluation on several samples with different surface conductivity is done with a new procedure based on insertion loss data obtained from a dual-TEM cell.

## II. SHIELDING EVALUATION

Attenuation of the electromagnetic energy may be characterised by the shielding effectiveness, SE and the insertion loss, IL. Screening effectiveness SE is defined as the ratio of electromagnetic field strengths $E_0/E_1$ measured without and with the tested material when it separates field source and receptor respectively: $SE_{dB} = 20 \log E_0/E_1$, in dB. The insertion loss IL is given by the attenuation in a transmitted signal caused by tested material insertion in measuring channel: $IL_{dB} = 10 \log U_0/U_1$, $U_0$ being the channel output voltage without the tested material and $U_1$ the same voltage with the tested material.

Methods for measuring shielding effectiveness in screened rooms are commonly used and subjected to standardisation [7-9]. The shielding effectiveness of base materials is, for instance, determined using the insertion-loss method, described in ASTM D4935.

Measurements of IL can be also based on the use of a dual-TEM cell [7,10] (see Fig.1). A typical TEM (transverse electromagnetic) cell consists of a section of rectangular coaxial transmission line tapered down at each end to match ordinary $50\,\Omega$ coaxial line. The TEM cell is well established as a device that creates a known broad-band isolated test field. A dual-TEM cell is then simply a pair of TEM cells with the added feature of an aperture in a shared wall. The aperture transfers power from the driving cell fed at the Port 1 to the receiving cell. The insertion loss provided by putting a sample on the aperture gives an evaluation of the shielding effectiveness of the material (tested at Port 2 or 4).

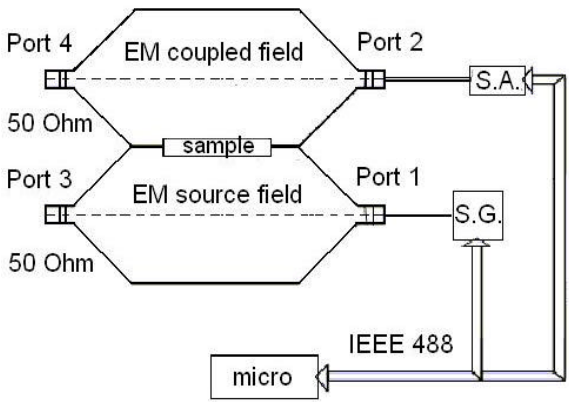

FIG.1: Dual-TEM cell for IL measurements. Samples are placed between the two cells. A signal generator creates the field in the lower cell and a signal analyser checks field in the upper cell. Both signals are sent to a microcomputer.

A TEM cell is primarily a low-frequency device where shielding effectiveness evaluation suffers from cell resonances. At resonance frequencies, the electrical field shows maxima and minima dependent on the experimental set-up (900 MHz for our device). Since TEM cells work at low frequency, they can be investigated with small-aperture theories. Casey [11] developed a solution modelling the sample as an equivalent impedance sheet, with

$$Z_S = 1/{\sigma h} \qquad (1)$$

where $\sigma$ is the sample conductivity and $h$ the thickness of the sheet. The impedance corresponds to the surface resistance of the sample.

Wilson and Ma [7] obtained for the dual-TEM cell, the following expression for the insertion loss, measuring the signal at Port 2:

$$IL_2 = L_S + 20\,log\left|1 + \frac{j4\omega\mu_o r \sigma h}{3\pi(1 + 2\pi R_C \sigma h)}\right| \qquad (2)$$

in dB, where $R_C$ is the contact resistance between the sheet under test and the cell aperture. $r$ is a factor coming from the aperture geometry. $L_S$ is a static constant with value 3.2 at Port 2, for high conducting samples. From this equation, it is easy to see that when a resistive material is used to load the aperture, the insertion loss is linearly dependent on frequency.

For calibration, we inserted on the aperture a sheet of Ni/Ag/Nylon nonwoven fabric, with a known surface resistance $R_S$, measured independently with a four-point set-up to be 0.09 $\Omega$.

In Eq.(2), factor $r$ depends on the shape and dimension $l$ of the dual- TEM aperture: it was estimated for a square shape as $0.579 \cdot l$ by Wilson and Ma [7]. The other two parameters, the contact resistance $R_C$ and the surface impedance $Z_S$, can be obtained with a best-fit in the frequency behaviour of the theoretical insertion $IL_{th}$ according to Eq.(2), with experimental data $IL_{exp}$ collected by means of a microcomputer connected with Port 2.

The best-fit is performed by minimizing the square deviation:

$$\Delta = \sum_{frequency}(IL_{th} - IL_{exp})^2 \qquad (3)$$

over all the tested frequencies. Since $IL_{th}$ is a function of $Z_S$ and $R_C$, the square deviation depends on them too, $\Delta = \Delta(Z_S, R_C)$. Searching $\Delta$ minima, we obtain the values of $Z_S$, $R_C$ giving the best agreement with experimental data.

For Ni/Ag/Nylon nonwoven fabric, the minimization procedure gives $Z_S = 0.084\,\Omega$, with $R_C = 0.02\,\Omega$. The theoretical curve corresponding to $Z_S = 0.084\,\Omega$ is given in Fig.2 (curve *a*), compared with the experimental data. If the insertion loss $IL_{th}$ for $0.084\,\Omega$ is evaluated neglecting the contact resistance, we obtain curve *b* in Fig.2.

Following Colaneri and Shacklette [12], to compare the results obtained with the dual-TEM cell with measurements obtained with other methods, the shielding effectiveness in far-field at low frequency is introduced as:

$$SE_{LF} = 20\,log\left(1 + \frac{Z_o/Z_S}{2}\right) \quad (4)$$

where $Z_0$ is the vacuum impedance. With Eq.(4), we have a $SE_{LF}$ ranging from 66.4 dB to 68. dB for $Z_S$ passing from 0.09 to 0.084 $\Omega$.

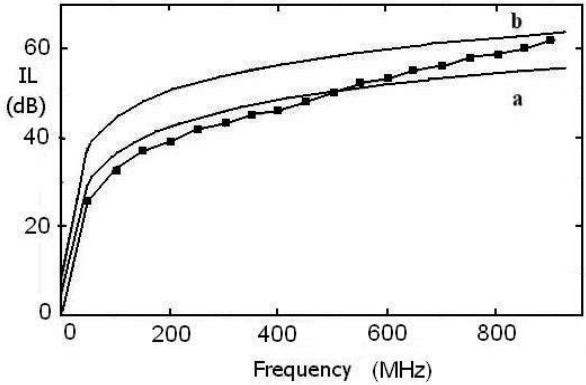

FIG.2: Insertion loss dual-TEM data for a Ni/Ag/Nylon nonwoven fabrics compared with theoretical $IL_{th}$ (Eq.2) with $Z_S = 0.09\,\Omega$, and $R_C = 0.02\,\Omega$ (curve a). Assuming a negligible contact resistance, $IL_{th}$ is shown by curve b.

## III. SAMPLES PREPARATION AND IL TESTS

Metals or metal- coated materials generally show very high EMI-SE, as we have seen in the previous section. However, they cannot be used as electromagnetic wave absorber since high conductivity makes them shield by surface reflection. Innovative materials, such as intrinsically conducting polymers (ICPs) are able to absorb as well as reflect electromagnetic waves, and then can exhibit certain advantages over metallic materials.

ICPs are conjugated polymers, with alternating single and double bonds in the polymer backbone, a necessary condition for charge carriers to move freely along the chain when doping is provided. The most prominent ICPs in EMI-SE are polypyrrole and polyaniline, where electrical conductivity can have values comparable to those observed for poorly conducting metals and alloy. ICPs do not require conductive fillers in order to provide shielding, so they may be used with or without fillers. In the presence of a conductive filler, an electrically conducting polymer matrix has the added advantage of being able to electrically connect filler units that do not touch one another, thereby enhancing the connectivity.

One of the first commercial textile products incorporating conductive polypyrrole was the Contex® conductive textile product line. More recently, textiles with a modified PPy coating have been commercially developed that are more conductive and thermally stable. Almost all fabrics can be coated using the aqueous process.

While imparting electrical conductivity and a dark colour to the substrates, the coating process barely affects the strength, drape, flexibility, and porosity of the starting substrates.

For the measurements of shielding effectiveness, polypyrrole coated fabrics were prepared similarly as described previously in [13,14], with raw chemicals purchased from Sigma-Aldrich and used without further purification. Stochiometric molar ratio of organic acid dopant, anthraquinone-2-sulfonic acid to pyrrole-monomer (i.e., 0.33:1) was used to ensure complete doping level. The molar ratio of polymerisation catalyst, iron(III) nitrate, to monomer (pyrrole) equal to 2.3 mol/mol was used for all reactions.

Simultaneous *in-situ* polymerisation and deposition of conductive polypyrrole leads to production of conductive, smooth and uniform coating with thickness under 1 micron, according to transmission electronic microscope measurements (see Fig.3).

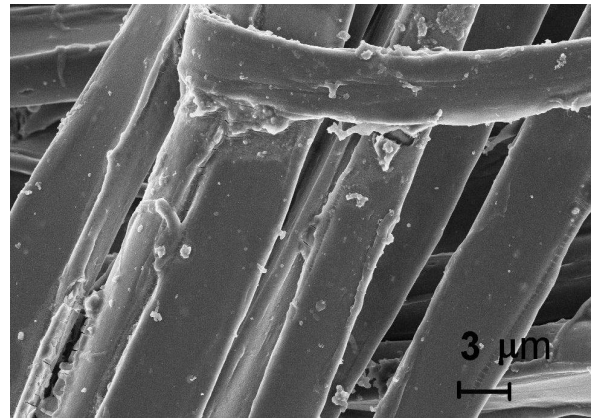

FIG.3: Scanning electron microscope images of one of the samples with polypyrrole coated fibers.

The electrical DC surface resistance $R_S$ was measured by using a four-in-line point probe in combination with computerised Loresta-AP meter from Mitsubishi Petrochemical Co., LTD.

Several samples of PPy-coated polyester nonwoven, ranging from 3 to 40 $\Omega/sq$ DC surface resistance were prepared. With the dual-TEM we obtain the insertion losses IL reported in Fig.4. In the figure, we can easily see that an increase in IL is obtained for high surface conductivity. A better conductivity produces reflection of waves, and the behaviour of the shield is going to be like that of metals.

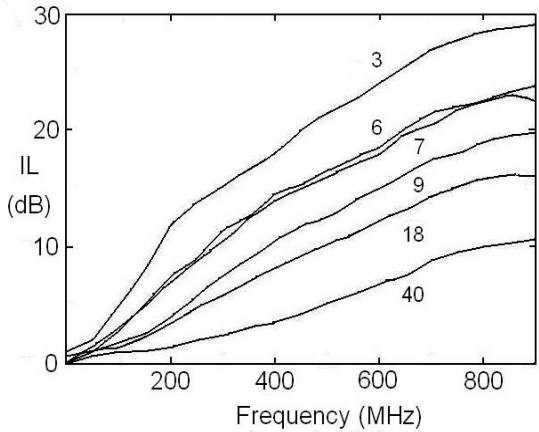

FIG.4: IL data obtained with the dual-TEM cell for several samples of polypyrrole coated fabrics with different nominal DC surface resistances ranging from 3 to 40 $\Omega/sq$.

On these data we apply the minimization procedure tested on the metallized sample, to have the effective surface impedance $Z_S$ : we obtain $Z_S$ corresponding to 2.7, 5, 6.5, 8.1, 15.5 and 32. $\Omega/sq$. For all the samples, we find $R_C$ negligible. According to Eq.(4), the shielding effectiveness at low frequency evaluated with $Z_S$ are 37.2, 31.8, 29.1, 27.0, 21.7 and 15.4 dB respectively. The static constant $L_S$ turns out to be negligible too. The DC surface resistance $R_S$ was determined with an uncertainty of about 10%; the two samples with 6. and 7. $\Omega/sq$ have then essentially the same surface impedance $Z_S$ and the same behaviour in IL.

In Fig.5, curves (a),(b) represent the theoretical insertion losses obtained with the best-fit for samples with nominal surface resistance $R_S$ equal to 18 $\Omega$ and 40 $\Omega$: from the best-fit with experimental data, we have $Z_S$ = 15.5 $\Omega$ and 32 $\Omega$ respectively.

Our observations are in good agreement with other published results on polypyrrole/PET woven fabric (Kim et al., [15]). The specific volume resistivity of composites prepared by Kim et al. was extremely low as 0.2 $\Omega.cm$ and SE was of about 36 dB over a wide frequency range up to 1.5 GHz. EMI-SE gradually increased from 13 to 26 dB with decrease of the specific volume resistivity in the region from 2.85 to 0.75 $\Omega.cm$ and then more steeply increased to 36 dB below 0.75 $\Omega.cm$, which is due to increase of the conductivity toward a metallic conductivity. The relationship between SE and electrical conductivity of the PPy composite obtained in [14] coincides with the relationships reported for other conducting polymer systems such as polyaniline film [16], polyaniline/ polymer composite [17].

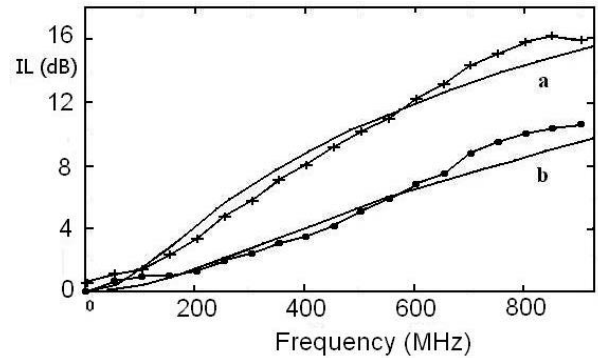

FIG.5: Curve (a) gives the best-fit for a sample with nominal DC surface resistance $R_S$ of 18 $\Omega/sq$: minimization procedure gives $Z_S$ =15.5 $\Omega/sq$ and $R_C$ = 0.$\Omega$ . Curve (b) represents the best-fit for the sample with a nominal DC surface resistance of 40 $\Omega/sq$: the minimization parameters are $Z_S$ = 32. $\Omega/sq$, $R_C$ = 0. $\Omega$ .

To evaluate the role of reflection and absorption in the shielding effectiveness of these textiles, we report in Table I, the insertion loss measured with the dual-TEM cell at 800 MHz, the shielding effectiveness according to Eq.4, reflection $R$ and absorption coefficient $A$, given according to the following equations:

$$R = 20 \log\left(\frac{(Z_0 + Z_S)^2}{4 Z_0 Z_S}\right) \quad (5)$$

$$A = 8.68 \frac{h}{d} \quad (6)$$

where $d$ is the skin depth of the sample. $h$ is the thickness of the shield. In the case of a metallized fabric, $h$ is the thickness of the coating, approximately $2\mu m$. For the non-woven fabrics with polypyrrole, the thickness we use for evaluating absorption coefficient is the thickness of the sample. The skin depth is given by $d = (\pi\nu\mu\sigma)^{-1/2}$, where $\nu$ is the frequency, $\mu$ the magnetic permeability, and $\sigma = 1/Z_S h$ the bulk conductivity. The relative permeability is 1 for these fabrics. Since the skin depth is depending on the frequency, we evaluated it at 800 MHz. The sample thickness is the same for all the samples and is *0.4 mm*.

Assuming the re-reflection contribution, $B$, can be determined as

$$B = IL - R - A, \quad (7)$$

from Table I, $B$ seems to drop off as the resistance increases. This is consistent with the skin depth increasing for less conductive materials.

These re-reflection contributions are probably an overstatement of the true values, due to the structure of non-woven fabrics. These materials are built with multiply stacked dielectric fibers with conductive coating layers around them, typically on the order of tenths of microns thick. This structure would lead to other mechanisms of shielding besides just simple, planar reflections and absorptions, which reduce the re-reflection term. The actual re-reflection gain is probably only a few dBs.

Eq. 4 consistently and substantially overestimates the shielding effectiveness compared with measured *IL*; in fact, it formally applies to low-frequency electric far-fields and does not account for re-reflections. As reported in Ref.18, $B$ can be neglected for electric fields and plane waves; the re-reflection contribution to *IL* could correspond in a dual-TEM to a sizeable H-field component.

As shown in Table I, the absorption of the textiles with polypyrrole is relevant in the samples with low surface impedance. Absorption $A$ constitutes the 20% of the insertion loss, for the sample with the surface impedance of $3. \Omega/sq$, and 15% for the sample with $18. \Omega/sq$. The absorption of metallized fabric is only 4 %.

Table I. SE, *IL*, $R$, $A$ and $B$ (see text for details)

| $R_S$ ($\Omega$) | $Z_S$ ($\Omega$) | SE (dB) | IL (dB) | R (dB) | A (dB) | B (dB) |
|---|---|---|---|---|---|---|
| 0.09 | 0.08 | 67.04 | 59. | 61.02 | 2.47 | -4.49 |
| 3 | 2.7 | 37.02 | 29. | 30.98 | 5.94 | -7.92 |
| 6 | 5. | 31.78 | 22. | 25.75 | 4.36 | -8.11 |
| 7 | 6.5 | 29.56 | 22. | 23.54 | 3.82 | -5.36 |
| 9 | 8.1 | 27.72 | 19. | 21.70 | 3.42 | -6.12 |
| 18 | 15.5 | 22.41 | 16.5 | 16.40 | 2.47 | -2.37 |
| 40 | 32. | 16.80 | 10. | 10.81 | 1.71 | -2.52 |

## IV. CONCLUSIONS

By means of dual-TEM measurements we find an EMI-SE of 37 dB for new modified PPy coated fabrics: for these materials, it is comparable to or even higher than EMI-SE reported for polyaniline systems or metal-coated carbon fiber composites. Increase of EMI-SE with the electrical conductivity results from the increase in shielding by reflection, due to the decrease of surface resistivity. A high reflection coefficient is due to a shallower skin depth of the composite with higher electrical conductivity. The PPy coated fabrics then, for EMI suppression related applications, have the advantage of a relative shielding efficiency that can be controlled by changing the surface electric conductivity.

Another important new result is the high absorption coefficient, displayed by the PPy non-woven fabrics, compared with that of a metallized fabric. We observed that absorption has a considerable share (up to 20%) of the total shielding effect. The shielding efficiency of metallized textile fabrics mainly derives from energy reflection, and not from its absorption. In many cases, such a phenomenon cannot be considered as a good one. PPy fabrics confirm their greater capability of absorbing electromagnetic radiation.

## V. AKNOWLEDGEMENT

The authors are indebted with Angelica Chiodoni for the SEM analysis.